\documentclass[10pt,twocolumn,prl,aps,amssymb,amsmath,tightenlines,showpacs]{revtex4}

\usepackage[dvips]{graphicx}
\usepackage{dsfont}
\newcommand{\re}{{\rm e}}
\newcommand{\ri}{{\rm i}}

\begin{document}
\title{Consistency of PT-symmetric quantum mechanics}
\author{Dorje~C.~Brody}%${}^2$, Hugh~F.~Jones${}^3$, 
%Bernhard~K.~Meister}

\affiliation{Department of Mathematics, Brunel University London, Uxbridge UB8 3PH, United Kingdom, and  \\ 
Department of Optical Physics and Modern Natural Science, 
St Petersburg National Research University of Information Technologies, Mechanics and Optics, 
49 Kronverksky Avenue, St Petersburg 197101, Russia} 
\begin{abstract}
In recent reports, suggestions have been put forward to the effect that parity and time-reversal (PT) symmetry in quantum mechanics is incompatible with causality. It is shown here, in contrast, that PT-symmetric quantum mechanics is fully consistent with standard quantum mechanics. This follows from the surprising fact that the much-discussed \textit{metric operator} on Hilbert space is not physically observable. In particular, for closed quantum systems in finite dimensions there is no statistical test that one can perform on the outcomes of measurements to determine whether the Hamiltonian is Hermitian in the conventional sense, or PT-symmetric---the two theories are indistinguishable. Nontrivial physical effects arising as a consequence of  PT symmetry are expected to be observed, nevertheless, for open quantum systems with balanced gain and loss. 
\end{abstract}

 
\maketitle

Since the realisation by Bender and Boettcher \cite{BB} that complex---non-Hermitian---Hamiltonians 
admitting space-time reflection (parity and time-reversal) symmetry can possess entirely real eigenvalues, 
considerable amount of research has been carried out into identifying properties of physical 
systems described by PT-symmetric Hamiltonians. It was subsequently observed that such Hamiltonians, 
although not Hermitian, can nevertheless be used to generate unitary time evolutions for the characterisation 
of closed quantum systems, provided that one works with a Hilbert space equipped with a preferentially 
selected 
inner product \cite{BBJ,AM}. In fact, the idea of modifying the Hilbert space inner product in quantum 
mechanics has previously been proposed in \cite{SGH}, but the work of \cite{BBJ,AM} has triggered 
extensive research into the identification of appropriate inner products---the so-called metric operators---for 
a wide range of complex PT-symmetric Hamiltonians. Nevertheless, the physical significance of the metric 
operator has hitherto remained elusive, leading to a variety of controversial claims concerning what might 
be achievable by altering the inner product in a laboratory experiment. 

The purpose of the present paper is to unambiguously settle this issue by showing that the degrees of 
freedom in the Hamiltonian associated with the choice of the Hilbert space inner product are not observable. 
Putting the matter differently, the lack of Hermiticity of the Hamiltonian, or equivalently the Petermann factor 
(see, e.g., \cite{Berry}), is not observable in closed systems---in contrast to open systems. The 
significance of this result is that not all parametric degrees of freedom in a complex Hamiltonian can be 
perturbed by an experimentalist in a laboratory. Our finding 
thus answers the open question raised in \cite{bbjm} regarding the possible physical constraints that prohibit 
experimentalists modifying the inner product (or equivalently, 
switching between different PT-symmetric Hamiltonians) in a 
laboratory. For the same token, the results here invalidate the claim in \cite{lee} that local PT-symmetric 
Hamiltonians violate the no-signalling condition, and the claim in \cite{CCC} that local PT-symmetric 
Hamiltonians can be used to increase entanglement. It will be shown that, 
on the contrary, there are no experiments that can be 
performed to a system modelled on a finite-dimensional Hilbert space whose statistics can distinguish 
between Hermitian and complex PT-symmetric Hamiltonians. For open quantum systems, on the other hand, 
for which a PT-symmetric environment can artificially be created, we expect interesting new physics to 
emerge owing to the presence of complex degeneracies. We conclude the paper with brief remarks on what 
one might expect to observe in such open quantum systems. 

To proceed we find it convenient to adopt the biorthogonal framework to describe properties of PT-symmetric 
Hamiltonians (as in \cite{AM3,Curtright,Curtright2,mannheim,DCB}). The use of biorthogonal states is of course 
common, and has 
been proven to be highly effective, in the literature of resonance physics \cite{IR}. Following closely the 
formalism of \cite{DCB}, we shall find it equally effective in the analysis of closed quantum 
systems. %, although the physical interpretation of some of the quantities are different. 
Here we take a `bottom up' approach by starting with a set of linearly independent vectors 
$\{|\phi_n\rangle\}_{n=1,\ldots,N}$ in an 
$N$-dimensional Hilbert space. We have in mind the case in which these vectors are not necessarily orthogonal. 
%, although they can be orthogonal, in which case there is a family of Hermitian operators whose eigenstates 
%correspond to $\{|\phi_n\rangle\}$. 
For a given $|\phi_k\rangle$, let us write $|\chi_k\rangle$ for an element of the 
one-dimensional subspace of the Hilbert space orthogonal to the span of $\{|\phi_n\rangle\}_{n\neq k}$. In this 
manner, we obtain a set of biorthogonal states: $\langle\chi_n|\phi_m\rangle \propto \delta_{nm}$, 
%\begin{eqnarray}
%\langle\chi_n|\phi_m\rangle \propto \delta_{nm} 
%\end{eqnarray}
which is unique up to scales. Since the vectors $\{|\phi_n\rangle\}$ are linearly independent, an arbitrary given 
state $|\psi\rangle$ can be expanded uniquely in the form 
\begin{eqnarray}
|\psi\rangle = \sum_{n} c_n |\phi_n\rangle . 
\label{eq:2} 
\end{eqnarray}
We then define a state $|{\tilde\psi}\rangle$ associated to $|\psi\rangle$ by the prescription 
\begin{eqnarray}
|{\tilde\psi}\rangle = \sum_{n} c_n |\chi_n\rangle . 
\label{eq:3} 
\end{eqnarray}
The term `associated state' was introduced in \cite{Pell} in the context of real Hilbert spaces. With the introduction 
of the notion of an associated state, we define a physical Hilbert space according to the following scheme: If 
$|\psi\rangle = \sum_n c_n |\phi_n\rangle$ and $|\varphi\rangle = \sum_n d_n |\phi_n\rangle$, then 
\begin{eqnarray}
\langle{\varphi},\psi\rangle \equiv 
\langle{\tilde\varphi}|\psi\rangle = \sum_{n,m} {\bar d}_n c_m \langle\chi_n|\phi_m \rangle 
= \sum_{n} {\bar d}_n c_n . 
\label{eq:HS} 
\end{eqnarray} 
Here we have fixed without loss of generality the normalisation convention so that $\langle\chi_n|\phi_m\rangle 
= \delta_{nm}$. With this norm convention, an operator acting on the states of the physical Hilbert space can 
be expressed in the form 
\begin{eqnarray}
{\hat F} = \sum_{n,m} f_{nm} |\phi_n\rangle\langle\chi_m| ,
\label{eq:5}
\end{eqnarray} 
whose expectation in a generic state (\ref{eq:2}) is 
\begin{eqnarray}
\langle{\hat F}\rangle = \frac{\langle{\tilde\psi}|{\hat F}|\psi\rangle}{\langle{\tilde\psi}|\psi\rangle} = 
\frac{\sum_{n,m} {\bar c}_n c_m f_{nm}}{\sum_n {\bar c}_n c_n} . 
\label{eq:z54} 
\end{eqnarray} 
For an operator ${\hat F}$ to represent a physical observable, we require its expectation values to be real. 
Evidently, this is realised only if the array $\{f_{nm}\}$ satisfies the `Hermiticity' condition 
${\bar f}_{mn}=f_{nm}$, even though ${\hat F}$ is not Hermitian: ${\hat F}^\dagger \neq {\hat F}$. This leads 
to the important observation that the reality condition, required for 
physical observables, merely demands the Hermiticity of the array $\{f_{nm}\}$, with respect to \textit{any} 
choice of biorthogonal basis states, but not the Hermiticity of the operator ${\hat F}$. 

Now suppose that ${\hat F}$ is a physical observable, that is, the array $\{f_{nm}\}$ is Hermitian with respect to 
a choice of biorthogonal basis, and that the system is prepared in a state $|\psi\rangle= \sum_n c_n 
|\phi_n\rangle$. Write $\{f_n\}$ for the eigenvalues of ${\hat F}$, which by definition are real, and $\{|f_n\rangle\}$ 
for the associated eigenstates. Then the probability of finding the outcome $f_k$ from the measurement of 
${\hat F}$, when the system is in the state $|\psi\rangle$, is given by 
\begin{eqnarray}
p_k = \frac{\langle{\tilde f}_k|\psi\rangle \langle{\tilde\psi}|f_k\rangle}
{\langle{\tilde\psi}|\psi\rangle\langle {\tilde f}_k|f_k\rangle} .
\label{eq:q18}
\end{eqnarray} 
It can easily be verified that $p_n\geq 0$ and that $\sum_n p_n=1$, even though $\langle f_n|f_m\rangle \neq 0$. 
We thus observe that a consistent probabilistic interpretation of quantum mechanics emerges, once again, from 
\textit{any} choice of complete 
biorthogonal basis states, irrespective of the numerical values of $|\langle f_n|f_m\rangle| 
= |\langle \phi_n|\phi_m\rangle|$.  

As regards the dynamics, suppose that ${\hat H}$ is the Hamiltonian of a system, with eigenstates 
$\{|\phi_n\rangle\}$ and real eigenvalues $\{E_n\}$. Since ${\hat H}^\dagger\neq{\hat H}$, the evolution 
operator ${\hat U} = \re^{-{\rm i}{\hat H}t/\hbar}$ does not satisfy the conventional 
unitarity condition: ${\hat U}^\dagger{\hat U}\neq{\mathds 1}$. 
Nevertheless, the time evolution \textit{is} unitary in the physical Hilbert space, which can easily be verified. 
Specifically, the solution to the Schr\"odinger equation $\ri \hbar \partial_t |\psi_t\rangle = {\hat K}|\psi_t\rangle$, 
%\begin{eqnarray}
%\ri \hbar \partial_t |\psi_t\rangle = {\hat K}|\psi_t\rangle,
%\label{eq:q54} 
%\end{eqnarray}
with initial condition $|\psi_0\rangle=\sum_nc_n|\phi_n\rangle$, is given by $|\psi_t\rangle = \sum_n c_n \re^{-{\rm i}E_n t/\hbar} |\phi_n\rangle$. 
%\begin{eqnarray}
%|\psi_t\rangle = \sum_n c_n \re^{-{\rm i}E_n t/\hbar} |\phi_n\rangle.
%\end{eqnarray}
According to the conjugation rule (\ref{eq:3}) we thus deduce that  
$\langle{\tilde\psi_t}|\psi_t\rangle = \langle{\tilde\psi_0}|\psi_0\rangle$ for all $t>0$. More generally, 
if $|\varphi_t\rangle$ is another solution to the Schr\"odinger equation with a different initial condition, 
then we have the unitarity condition 
$\langle{\tilde\varphi_t}|\psi_t\rangle = \langle{\tilde\varphi_0}|\psi_0\rangle$ for all $t>0$, on account 
of the reality of $\{E_n\}$. 

The foregoing analysis shows that based on the biorthogonal formalism, the reality of physical observables, the 
consistency of probabilistic interpretation of measurements, and the unitarity of the time evolution all follow 
straightforwardly, without the knowledge of the numbers $\langle \phi_n|\phi_m\rangle$. These observations 
suggest that for a closed quantum system, the numerical values of the overlaps $\langle \phi_n|\phi_m\rangle$ 
may not be physically identifiable. We shall proceed to establish that this is indeed the case. Our proof is based 
on the construction of physical observables. Namely, given the biorthogonal basis states $\{|\phi_n\rangle, 
|\chi_n\rangle\}$, we can construct the algebra of all physical observables directly, without reference to any other 
ingredients. Then an arbitrary physical observable, up to trace, is given by a linear combination of the elements 
of the algebra. The parameters entering in the algebra, which are related to the numbers 
$\langle \phi_n|\phi_m\rangle$, however, are, by construction, completely hidden to an 
observer, since expectation values only depend on the state and the expansion coefficients of the observable in 
terms of the elements of the 
algebra. This completes the proof that the numbers $\langle \phi_n|\phi_m\rangle$ cannot be observed or 
estimated from any measurement in a closed quantum system. It follows, in particular, that the so-called metric 
operator, of which an enormous amount of research efforts have been made in the literature, carries no physical 
information, again in the case of a closed system. 

To gain a better intuition of the proof above, it suffices to consider the two level system. We let 
$\{|e_1\rangle,|e_2\rangle\}$ be a pair of orthonormal states, and construct an arbitrary pair of 
linearly independent vectors. Without loss of generality we can set $|\phi_1\rangle\propto 
|e_1\rangle$ and $|\phi_2\rangle\propto \cos\frac{1}{2}\xi|e_1\rangle+ \sin\frac{1}{2}\xi 
\re^{{\rm i}\eta}|e_2\rangle$; but to make the resulting algebra looking a little more 
symmetric let us choose the parameterisation as follows: Fixing the normalisation such 
that $\langle\chi_n|\phi_m\rangle=\delta_{nm}$, we set 
\begin{eqnarray} 
\left\{ \begin{array}{l} 
|\phi_1\rangle = \frac{1}{\sqrt{2\sin\frac{1}{2}\xi}} \left[ (\cos\frac{1}{4}\xi+\sin\frac{1}{4}\xi) 
|e_1\rangle \right. \\ \left. \qquad \qquad \qquad \qquad + (\cos\frac{1}{4}\xi-\sin\frac{1}{4}\xi) 
\re^{{\rm i}\eta}  |e_2\rangle \right]
\\ 
|\phi_2\rangle = \frac{1}{\sqrt{2\sin\frac{1}{2}\xi}} \left[ (\cos\frac{1}{4}\xi-\sin\frac{1}{4}\xi) 
|e_1\rangle \right. \\ \left. \qquad \qquad \qquad \qquad + (\cos\frac{1}{4}\xi+\sin\frac{1}{4}\xi) 
\re^{{\rm i}\eta}  |e_2\rangle \right]
\end{array} \right. 
\label{eq:10} 
\end{eqnarray}
and 
\begin{eqnarray} 
\left\{ \begin{array}{l} 
|\chi_1\rangle = \frac{1}{\sqrt{2\sin\frac{1}{2}\xi}} \left[ (\cos\frac{1}{4}\xi+\sin\frac{1}{4}\xi) 
|e_1\rangle \right. \\ \left. \qquad \qquad \qquad \qquad - (\cos\frac{1}{4}\xi-\sin\frac{1}{4}\xi) 
\re^{{\rm i}\eta}  |e_2\rangle \right]
\\ 
|\chi_2\rangle = \frac{-1}{\sqrt{2\sin\frac{1}{2}\xi}} \left[ (\cos\frac{1}{4}\xi-\sin\frac{1}{4}\xi) 
|e_1\rangle \right. \\ \left. \qquad \qquad \qquad \qquad - (\cos\frac{1}{4}\xi+\sin\frac{1}{4}\xi) 
\re^{{\rm i}\eta}  |e_2\rangle \right]
\end{array} \right. 
\label{eq:11} 
\end{eqnarray}
for an arbitrary biorthogonal basis states, parameterised by $(\xi,\eta)$. 
(Here we have chosen the normalisation such that $\langle\phi_1|\phi_1\rangle= 
\langle\phi_2|\phi_2\rangle=\langle\chi_1|\chi_1\rangle=\langle\chi_2|\chi_2\rangle$, so that all 
the states are defined on the same projective Hilbert space---this removes the scale ambiguity 
discussed for example in \cite{Znojil,LVJ,IM}.)
In the case of two-level systems, a 
physical observable, up to trace, can be expressed as a linear combination of the three Pauli matrices, 
with real coefficients. We thus use (\ref{eq:10}) and (\ref{eq:11}) directly 
to work out the three Pauli matrices: 
\begin{eqnarray}
{\hat\sigma}_x = \left( \begin{array}{cc} 0 & \re^{-{\rm i}\eta} \\ 
\re^{{\rm i}\eta} & 0 \end{array} \right) , 
\end{eqnarray} 
\begin{eqnarray}
{\hat\sigma}_y = \left( \begin{array}{cc} \ri \cot\frac{1}{2}\xi &  -\ri 
\csc\frac{1}{2}\xi \re^{-{\rm i}\eta} \\ 
\ri \csc\frac{1}{2}\xi \, \re^{{\rm i}\eta} & - \ri \cot\frac{1}{2}\xi \end{array} \right) ,
\end{eqnarray} 
and 
\begin{eqnarray}
{\hat\sigma}_z = \left( \begin{array}{cc}  \csc\frac{1}{2}\xi  &  - \cot\frac{1}{2}\xi \, \re^{-{\rm i}\eta} \\ 
\cot\frac{1}{2}\xi \, \re^{-{\rm i}\eta} & - \csc\frac{1}{2}\xi \end{array} \right) . 
\end{eqnarray} 
It should be evident by a straightforward computation that the triplet 
$({\hat\sigma}_x,{\hat\sigma}_y, {\hat\sigma}_z)$ 
fulfils the standard $\mathfrak{su}(2)$ commutation relations. These extended Pauli matrices are in general 
not Hermitian (in this example ${\hat\sigma}_x$ is Hermitian), 
but their eigenvalues are nevertheless real and are given by $\pm1$. In fact, up to unitarity, these are the most 
general parametric family of Pauli matrices, while the standard Hermitian limit is restored if we let $\xi\to\pi$ and 
$\eta\to0$. The 
expectation values, in the sense of (\ref{eq:z54}), of these Pauli matrices in a generic state 
$|\psi\rangle = \cos\frac{1}{2}\theta|\phi_1\rangle + \sin\frac{1}{2}\theta\re^{{\rm i}\varphi}|\phi_2\rangle$ are 
then given by 
\begin{eqnarray}
\langle{\hat\sigma}_x\rangle=\sin\theta\cos\varphi, ~ 
\langle{\hat\sigma}_y\rangle=\sin\theta\sin\varphi, ~ 
\langle{\hat\sigma}_z\rangle=\cos\theta , 
\label{eq:15}
\end{eqnarray}
independent of the values of the parameters $(\xi,\eta)$. An arbitrary physical observable 
${\hat F}\neq {\hat F}^\dagger$ can thus 
be expressed in the form 
\begin{eqnarray}
{\hat F} = t {\mathds 1} + x {\hat\sigma}_x + y {\hat\sigma}_y + z {\hat\sigma}_z  , 
\label{eq:16} 
\end{eqnarray} 
whose expectation value in any given state (pure or mixed) is likewise independent of the 
parameters $(\xi,\eta)$. Note that the sextet $(t,x,y,z,\xi,\eta)$ corresponds to the six parameters 
required for specifying most general PT-symmetric $2\times2$ Hamiltonians \cite{wang,EMG}. 

The foregoing example illustrates the fact that when a Hamiltonian ${\hat H}$ is prescribed according to 
(\ref{eq:16}), the only parameters that an experimentalist can adjust in a laboratory are the ones in the 
triplet $(x,y,z)$, 
and possibly $t$ in some circumstances, but not the remaining two variables $(\xi,\eta)$. For the same 
token, the metric operator widely investigated in the literature is not physically observable. To 
see this, let us return to the general $N$-dimensional case. With respect to any given choice of orthonormal 
basis $\{|e_n\rangle\}_{n=1,\ldots,N}$, let us define an operator ${\hat u}$ by the property that $|\phi_n\rangle 
= {\hat u}|e_n\rangle$ for all $n$, and similarly define ${\hat v}$ by the relation $|\chi_n\rangle = {\hat v} 
|e_n\rangle$ for all $n$. Evidently, in the two-dimensional case the operators ${\hat u}$ and ${\hat v}$ are 
fully determined by the parameter pair $(\xi,\eta)$---they are essentially transpositions of the matrix of 
coefficients in (\ref{eq:10}) and (\ref{eq:11}).  
Then the linear independence of $\{|\phi_n\rangle\}$ implies that ${\hat u}$ is invertible, and the 
orthonormality condition $\langle\chi_n|\phi_m\rangle=\delta_{nm}$ implies that the inverse of 
${\hat u}$ is given by ${\hat v}^{\dagger}$, i.e. ${\hat v}^\dagger{\hat u}={\mathds 1}$. 

To proceed, let us substitute $|\phi_n\rangle={\hat u}|e_n\rangle$ and $|\chi_n\rangle = 
({\hat u}^\dagger)^{-1} |e_n\rangle$ in (\ref{eq:5}) for a physical observable ${\hat F}$ and obtain: 
\begin{eqnarray}
{\hat F} &=& {\hat u}\, {\hat f}\, {\hat u}^{-1},
\label{eq:17}
\end{eqnarray}
where 
\begin{eqnarray}
{\hat f} = \sum_{n,m} f_{nm} |e_n\rangle\langle e_m| 
\end{eqnarray}
is a standard Hermitian operator. The relation (\ref{eq:17}) thus defines a similarity transformation between a 
Hermitian operator ${\hat f}$ and a PT-symmetric operator ${\hat F}$. The operator ${\hat u}$ defining the 
transformation is unique up to unitary transformations associated with the choice of $\{|e_n\rangle\}$. We also 
observe that by taking the Hermitian conjugate of ${\hat F}$ we obtain ${\hat F}^\dagger=({\hat u}^\dagger)^{-1} 
\, {\hat f}\, {\hat u}^\dagger$, but (\ref{eq:17}) implies that ${\hat f}={\hat u}^{-1}\,{\hat F}\,{\hat u}$ so we deduce 
that  
\begin{eqnarray}
{\hat F}^\dagger = ({\hat u}{\hat u}^\dagger)^{-1}\, {\hat F} \, ({\hat u}{\hat u}^\dagger) , 
\label{eq:19} 
\end{eqnarray}
where ${\hat u}{\hat u}^\dagger$ is an invertible positive Hermitian operator. This is the \textit{reality condition} 
that has to be satisfied by all physical observables. It is worth remarking that in finite-dimensional Hilbert space, 
the structure of quantum mechanics is not sufficient to identify the reality of states and observables. One has 
to augment the Hilbert space with the specification of a positive Hermitian quadratic form ${\hat g}=
({\hat u}{\hat u}^\dagger)^{-1}$. Only then can one speak of the reality condition. The fact that Dirac had made 
the most convenient and economical choice ${\hat g}={\mathds 1}$ perhaps 
has had the effect of obscuring this subtlety, 
but it can be made fully transparent by adopting the biorthogonal framework. Finally, since $|\chi_n\rangle = 
({\hat u}^\dagger)^{-1} |e_n\rangle = ({\hat u}^\dagger)^{-1} {\hat u}^{-1}|\phi_n\rangle = {\hat g}|\phi_n\rangle$, 
and since ${\hat g}^\dagger={\hat g}$, we deduce that 
\begin{eqnarray}
\frac{\langle{\tilde\psi}| {\hat F}|\psi\rangle}{\langle{\tilde\psi}|\psi\rangle} = 
\frac{\langle{\psi}|{\hat g}{\hat F}|\psi\rangle}{\langle\psi| {\hat g}|\psi\rangle}  . 
\end{eqnarray}
In other words, the operator ${\hat g}$ plays the role of a metric for the (original) Hilbert space. The degrees 
of freedom associated with the metric operator, however, are precisely those hidden variables required for the 
specification of the algebra of observables. It follows that the metric operator ${\hat g}$ is not a physically 
measurable quantity. 

In the conventional approach to PT-symmetric quantum mechanics, it is common, and indeed natural, to adopt 
a `top-down' approach in which one begins with the specification of the Hamiltonian. As a consequence, the 
physical degrees of freedom and those associated with the specification of the algebra of observables are 
intertwined, and there is no elementary way of disentangling them. This has had the effect of obscuring the 
meaning and the significance of the metric operator. In the literature it has long been acknowledged, since 
Mostafazadeh \cite{AM2}, that (at least in finite dimensions) a PT-symmetric Hamiltonian, when its eigenstates 
are also PT-symmetric, is \textit{equivalent} to a Hermitian Hamiltonian via the similarity transformation 
(\ref{eq:17}). This conclusion, whilst technically correct, is perhaps misguided since it does not explain the 
physical significance of the transformation. Additionally, the equivalence statement ceases to hold in typical 
infinite-dimensional systems. The accurate statement that we can now make is that a PT-symmetric 
Hamiltonian, when its eigenstates are also PT-symmetric, is \textit{indistinguishable} from a Hermitian Hamiltonian. 
Furthermore, there are no physical operations one can perform to break PT symmetry in a closed system, since 
those operations require the manipulation of inaccessible variables. It also follows that the suggestion that 
PT-symmetry is a generalisation of Hermiticity \cite{mannheim}, in the case of finite-dimensional Hilbert spaces, 
is inaccurate, since the reality condition (\ref{eq:19}) must apply for 
some choice of ${\hat u}$ for all observables. 

Let us discuss further implications of the above findings. It was observed in \cite{bbjm} that if an experimentalist 
has a free access to alter the metric operator ${\hat g}$, then it is possible to transform one quantum state into 
another with finite energy in an arbitrary short time. The question as to whether there might be a physical constraint 
that limits this possibility was left open in \cite{bbjm}, but the above result clearly shows that indeed there is no way 
of altering 
${\hat g}$; hence an arbitrary fast state transformation cannot be implemented. An analogous conclusion was 
drawn in \cite{Croke}, however, not based on the observability of the metric operator, since the analysis therein 
is based on the incorrect assumption that the metric operator can be chosen or altered. In \cite{lee} it was argued 
that the application of a local PT-symmetric Hamiltonian can lead to the construction of a communication channel 
that violates the causality constraints. Again, their analysis, which assigns physical significance to the metric 
operator, is empty, and their conclusion that if a PT-symmetric Hamiltonian can ``coexist'' within the conventional 
quantum system then it cannot be regarded as describing viable physical system, since the similarity transformation 
to an equivalent Hermitian Hamiltonian requires nonlocal data, is false: Local observers do not have the capability 
to select the metric operator in a closed system. The correct treatment of a coupled system 
requires the consideration of the tensor product of two physical Hilbert spaces in the sense defined above. A 
similar analysis is presented in \cite{CCC} to suggest that the degree of entanglement can be modified by use of 
local PT-symmetric Hamiltonians---once again, this conclusion is obtained only by means of accessing the physically 
inaccessible parameters in the Hamiltonian. 

In summary, we have demonstrated that for the characterisation of a closed quantum system by means of a complex 
PT-symmetric Hamiltonian, not all the parameters in the Hamiltonian are experimentally accessible. This finding 
resolves several controversies and confusions in the literature. To what extent the foregoing analysis extends to 
infinite-dimensional systems is an interesting open question. It suffices to remark that in infinite dimensions, the 
biorthogonal partner $\{|\chi_n\rangle\}$ of a complete set of states $\{|\phi_n\rangle\}$ need not be complete 
\cite{KS}. The foregoing analysis based on the biorthogonal formalism nevertheless suggests the following 
conjecture: Provided that the Hamiltonian admits a complete biorthogonal set of basis states and a set of real 
eigenvalues, it can be used as a viable model for a physical system, irrespective of whether it admits an `equivalent'  
Hermitian counterpart Hamiltonian. In particular, the unboundedness of the metric operator or its inverse, which 
has shed doubts on the naive `equivalence' arguments of \cite{AM2,Croke}, will entail no implications. 

We conclude with a discussion on PT symmetry in open quantum systems. Since the observation 
that a PT-symmetric environment can be replicated by means of a balanced gain and loss \cite{RDM,DC1,NM}, 
various experimental realisations have either been predicted, or observed in laboratory experiments, for a 
variety of systems including optical waveguides \cite{DC2,DC3}, Bose-Einstein condensates \cite{graefe,wunner}, 
laser physics \cite{DS,SR,SR2}, electric circuits \cite{TK,TK2}, spectroscopy \cite{YNJ}, and microwave cavity 
\cite{BD,CMB}, to name a few. Unlike closed systems discussed above, the parametric freedom of the 
operator ${\hat u}$ in an open system \textit{can} be manipulated experimentally, and one can predict or observe 
counterintuitive behaviours arising from the degeneracies of the energy eigenstates (exceptional points). 
All the experiments that have been performed so far, however, involve classical systems admitting 
quantum-mechanical characterisations (e.g., paraxial approximation to the scalar Hermholtz equation in 
optics). 
The implementation of a controlled gain and loss in a \textit{bona fide} quantum system, on the other hand, 
seems to be achievable only probabilistically, owing to quantum limitations such as those arising from 
uncertainty principles. For instance, to maintain a perfect balance of gain and loss of energy, not only the 
amount of energy but also the timing of its flow will have to be controlled with a perfect precision. For classical 
systems this is not an issue, while quantum mechanically, gain and loss can be balanced at best only on average. 
One can nevertheless model the gain and loss 
channels by use of Lindblad operators to investigate the behaviour of such quantum systems. It seems evident, 
due to loss of information whenever gain and loss are implemented probabilistically, that the long-time limit of 
an open PT-symmetric quantum dynamics is such that any initial state will decay into the state of total ignorance. 
In view of the preliminary analysis of \cite{BG}, however, it is tempting to conjecture that the signature of 
exceptional points in the classical counterpart Hamiltonian is visible in the quantum dynamics: Namely, when the 
eigenvalues of the corresponding classical PT-symmetric Hamiltonian of the system are real, the decay of the 
quantum state into the state of total ignorance is superimposed with an oscillatory (unitary) motion, whereas in 
the region where PT-symmetry of the eigenstates are broken, the decay will be purely 
exponential, exhibiting no oscillation. Such a transition, if the stated 
conjecture were to hold, can be examined both theoretically and experimentally.

The author thanks participants of \textit{Quantum (and Classical) Physics with Non-Hermitian Operators}, 
Jerusalem, July 2015, for stimulating discussion and comments.

%\begin{references}


\begin{thebibliography}{999}

\vspace{-0.5cm}

\bibitem{BB}
C.~M.~Bender and S.~Boettcher  
``Real spectra in non-Hermitian Hamiltonians having PT-symmetry'' 
{\em Phys. Rev. Lett.} \textbf{80}, 5243--5246 (1998).

\bibitem{BBJ} 
C.~M.~Bender, D.~C.~Brody and H.~F.~Jones 
``Complex extension of quantum mechanics'' 
{\em Phys. Rev. Lett.} \textbf{89}, 27040 (2002). 

\bibitem{AM} 
A.~Mostafazadeh
``Pseudo-Hermiticity versus PT-symmetry III: 
Equivalence of pseudo-Hermiticity and the presence of antilinear symmetries'' 
{\em J. Math. Phys.} \textbf{43}, 3944--3951 (2002).

\bibitem{SGH} 
F.~G.~Scholtz, H.~B.~Geyer and F.~J.~W.~Hahne
``Quasi-Hermitian operators in quantum mechanics and the variational principle''  
{\em Ann. Phys.} \textbf{213}, 74--101 (1992).

\bibitem{Berry}
M.~V.~Berry 
``Mode degeneracies and the Petermann excess-noise factor for unstable lasers'' 
{\em J. Mod. Opt.} \textbf{50}, 63--81 (2003). 

\bibitem{bbjm} 
C.~M. Bender, D.~C. Brody, H.~F. Jones, and B.~K. Meister
Faster than Hermitian quantum mechanics. 
{\em Phys. Rev. Lett.} \textbf{98}, 040403 (2007). 

\bibitem{lee} 
Y.-C. Lee, M.-H. Hsiu, S.~T.~Flammia, and R.-K. Lee 
Local PT symmetry violates the no-signalling principle. 
{\em Phys. Rev. Lett.} \textbf{112}, 130404 (2014).

\bibitem{CCC} 
S.-L.~Chen, G.-Y.~Chen, and Y.-N.~Chen
Increase of entanglement by local PT-symmetric operations. 
{\em Phys. Rev.} A\textbf{90}, 054301 (2014). 

\bibitem{AM3} 
A.~Mostafazadeh 
``Pseudo-Hermiticity versus PT-symmetry. II. A complete
characterization of non-Hermitian Hamiltonians with a real spectrum'' 
{\em J. Math. Phys.} \textbf{43}, 2814--2816 (2002).

\bibitem{Curtright} 
T.~Curtright and L.~Mezincescu  
``Biorthogonal quantum systems'' 
{\em J. Math. Phys.} \textbf{48}, 092106 (2007). 

\bibitem{Curtright2} 
T.~Curtright, L.~Mezincescu, and D.~Schuster
``Supersymmetric biorthogonal quantum systems'' 
{\em J. Math. Phys.} \textbf{48}, 092108 (2007). 

\bibitem{mannheim}
P.~D.~Mannheim 
``PT symmetry as a necessary and sufficient condition for unitary time evolution'' 
{\em Phil. Trans. R. Soc.} A\textbf{371}, 20120060 (2013). 

\bibitem{DCB} 
D.~C.~Brody 
``Biorthogonal quantum mechanics'' 
{\em J. Phys.} A\textbf{47}, 035305 (2014). 

\bibitem{IR}
J.~Oko{\l}owicz, M.~P{\l}oszajczak and I.~Rotter  
``Dynamics of quantum systems embedded in a continuum'' 
{\em Phys. Rep.} \textbf{374}, 271--383 (2003).

\bibitem{Pell} 
A.~J.~Pell 
``Biorthogonal systems of functions''  
{\em Trans. Amer. Math. Soc.} \textbf{12}, 135--164 (1911). 

\bibitem{Znojil} 
M.~Znojil
``On the role of the normalization factors $\kappa_n$ and of the pseudo-metric 
${\cal P}\neq{\cal P}^\dagger$ in crypto-Hermitian quantum models'' 
{\em SIGMA} \textbf{4}, 001 (2008).

\bibitem{LVJ} 
A.~Leclerc, D.~Viennot and G.~Jolicard 
``The role of the geometric phases in adiabatic population tracking for 
non-Hermitian Hamiltonians'' 
{\em J. Phys.} A\textbf{45}, 415201 (2012).

\bibitem{IM} 
S.~Ib\'a${\tilde {\rm n}}$ez and J.~G.~Muga
``Adiabaticity condition for non-Hermitian Hamiltonians'' 
{\em Phys. Rev.} A\textbf{89}, 033403 (2014). 

\bibitem{wang} 
Q.~Wang, S.~Chia, and J.~Zhang 
``PT symmetry as a generalisation of Hermiticity'' 
{\em J. Phys.} A\textbf{43}, 295301 (2010).

\bibitem{EMG} 
E.~M.~Graefe, S.~Mudute-Ndumbe, and M.~Taylor 
``Random matrix ensembles for PT-symmetric systems'' 
{\em J. Phys.} A\textbf{48}, 38FT02 (2015).

\bibitem{AM2} 
A.~Mostafazadeh   
``Exact PT-symmetry is equivalent to Hermiticity'' 
{\em J. Phys.} A\textbf{36}, 7081--7091 (2003).

\bibitem{Croke} 
S.~Croke
``PT-symmetric Hamiltonians and their application in quantum information'' 
{\em Phys. Rev.} A\textbf{91}, 052113 (2015).

\bibitem{KS} 
S.~Kaczmarz and H.~Steinhaus 
{\em Theorie der Orthogonalreihen}, 2nd ed. 
(New York: Chelsea Publishing, 1951). 

\bibitem{RDM} 
A.~Ruschhaupt, F.~Delgado and J.~G.~Muga
``Physical realization of PT-symmetric potential scattering in a planar slab waveguide'' 
{\em J. Phys.} A\textbf{38}, L171--L176 (2005).

\bibitem{DC1} 
K.~G.~Makris, R.~El-Ganainy, D.~N.~Christodoulides, and Z.~H.~Musslimani 
``Beam dynamics in PT-symmetric optical lattices''
{\em Phys. Rev. Lett.} \textbf{100}, 103904 (2008).

\bibitem{NM} 
S.~Klaiman, U.~G\"unther, and N.~Moiseyev 
``Visualization of branch points in PT-symmetric waveguides'' 
{\em Phys. Rev. Lett.} \textbf{101}, 080402 (2008).

\bibitem{DC2} 
A.~Guo, G.~J.~Salamo, D.~Duchesne, R.~Morandotti, M.~Volatier-Ravat, 
V.~Aimez, G.~A.~Siviloglou, and D.~N.~Christodoulides 
``Observation of PT-symmetry breaking in complex optical potentials'' 
{\em Phys. Rev. Lett.} \textbf{103}, 093902 (2009).

\bibitem{DC3} 
C.~E.~R\"uter, K.~G.~Makris, R.~El-Ganainy, D.~N.~Christodoulides, and D.~Kip 
``Observation of parity-time symmetry in optics''
{\em Nat. Phys.} \textbf{6}, 192--195 (2010). 

\bibitem{graefe} 
E.~M.~Graefe 
``Stationary states of a PT symmetric two-mode Bose-Einstein condensate'' 
{\em J. Phys.} A\textbf{45}, 444015 (2012).

\bibitem{wunner} 
R.~Gut\"ohrlein, J.~Schnabel, I.~Iskandarov, H.~Cartarius, J.~Main and G.~Wunnern
``Realizing PT-symmetric BEC subsystems in closed hermitian systems'' 
{\em J. Phys.} A\textbf{48}, 335302 (2015). 

\bibitem{DS} 
Y.~D.~Chong, L.~Ge, and A.~D.~Stone 
``PT-symmetry breaking and laser-absorber modes in optical scattering systems'' 
{\em Phys. Rev. Lett.} \textbf{106}, 093902 (2011).

\bibitem{SR} 
L.~Ge, Y.~D.~Chong, S.~Rotter, H.~E.~T\"ureci, A.~D.~Stone 
``Unconventional modes in lasers with spatially varying gain and loss'' 
{\em Phys. Rev.} A\textbf{84}, 023820 (2011). 

\bibitem{SR2} 
M.~Liertzer, L.~Ge, A.~Cerjan, A.~D.~Stone, H.~T\"ureci, and S.~Rotter 
``Pump-induced exceptional points in lasers''  
{\em Phys. Rev. Lett.} \textbf{108}, 173901 (2012). 

\bibitem{TK} 
J.~Schindler, A.~Li, M.~C.~Zheng, F.~M.~Ellis, and T.~Kottos 
``Experimental study of active LRC circuits with PT symmetries'' 
{\em Phys. Rev.} A\textbf{84}, 040101 (2011).

\bibitem{TK2} 
H.~Ramezani, J.~Schindler, F.~M.~Ellis, U.~G\"unther, and T.~Kottos, 
``Bypassing the bandwidth theorem with PT symmetry'' 
{\em Phys. Rev.} A \textbf{85}, 062122 (2012). 

\bibitem{YNJ} 
Y.~N.~Joglekar, R.~Marathe, P. Durganandini and R.~K.~Pathak
``PT spectroscopy of the Rabi problem'' 
{\em Phys. Rev.} A\textbf{90}, 040101 (2014).

\bibitem{BD} 
S.~Bittner, B.~Dietz, U.~G\"unther, H.~L.~Harney, M.~Miski-Oglu, 
A.~Richter, and F.~Sch\"afer 
``PT symmetry and spontaneous symmetry breaking in a microwave billiard'' 
{\em Phys. Rev. Lett.} \textbf{108}, 024101 (2012). 

\bibitem{CMB} 
B.~Peng, S.~\"O.~Kaya, F.~Lei, F.~Minufu, M.~Gianfreda, G.~L.~Long, S.~Fan, 
F.~Nori, C.~M.~Bender and L.~Yang 
``Parity-time-symmetric whispering-gallery microcavities'' 
{\em Nat. Phys.} \textbf{10}, 394--398 (2014). 

\bibitem{BG} 
D.~C.~Brody and E.~M.~Graefe 
``Mixed-state evolution in the presence of gain and loss'' 
{\em Phys. Rev. Lett.} \textbf{109}, 230405 (2012). 


%\bibitem{bbcgms} C.~M. Bender, D.~C. Brody, J.~Caldeira, U.~G\"{u}nther, 
%B.~K. Meister, and B.~F. Samsonov
%{\em Phil. Trans. R. Soc. London} A\textbf{371}, 20120160 (2013).
%
%\bibitem{etc} 
%R.~Kretschmer and L.~Szymanowski 
%{\em Phy. Lett.} A\textbf{325}, 112 (2004); 
%F.~Bagarello 
%{\em J. Phy.} A\textbf{43}, 175203 (2010);  
%P.~Siegl and D.~Krej\v{c}i\v{r}\'{\i}k 
%{\em Phy. Rev.} D\textbf{86}, 121702(R) (2012); 
%C.~M.~Bender and S.~Kuzhel 
%{\em J. Phys.} A\textbf{43}, 444005 (2012); 
%A.~Mostafazadeh 
%{\em Phil. Trans. R. Soc. London} A\textbf{371}, 20120050 (2013). 
%
%\bibitem{bari} 
%N.~K.~Bari 
%{\em Moskov. Gos. Univ. U\v{e}nye Zapiski Matematika} 
%\textbf{148}, 69 (1951).

\end{thebibliography}
\end{document}